\documentclass[amssymb,aps,showpacs,floatfix,nofootinbib,showpacs,12pt]{revtex4}
\usepackage{graphicx}
\def\bwt{\begin{widetext}}

\def\ewt{\end{widetext}}

\def\be{\begin{equation}}

\def\ee{\end{equation}}

\def\bea{\begin{eqnarray}}

\def\eea{\end{eqnarray}}

\def\bean{\begin{eqnarray*}}

\def\eean{\end{eqnarray*}}

\def\bary{\begin{array}}

\def\eary{\end{array}}

\def\bit{\begin{itemize}}

\def\eit{\end{itemize}}

\def\su5u1{SU(5) \times U(1)}

\def\fsu5u1{SU(5) \times U(1)'}

\def\so10{SO(10)}

\def\sq20{SO(10) \times SO(10)}

\usepackage[centertags]{amsmath}

\usepackage{amssymb}

\begin{document}

\title{Non-Thermal Production of WIMPs, \\
Cosmic $e^\pm$ Excesses and $\gamma$-rays from the Galactic
Center}

\author{Xiao-Jun Bi$^{1,2}$, Robert Brandenberger$^{3,4,5,6,7}$,
Paolo Gondolo$^{8,6}$, Tianjun Li$^{9,10,6}$, Qiang Yuan$^{1}$ and
Xinmin Zhang$^{4,5,6}$}

\affiliation{ $^1$ Key Laboratory of Particle Astrophysics,
Institute of High Energy Physics, Chinese
Academy of Sciences, Beijing 100049, P. R. China \\
$^{2}$ Center for High Energy Physics, Peking University, Beijing
100871, P.R. China \\
$^{3}$ Department of Physics, McGill University, Montr\'eal, QC,
H3A 2T8, Canada \\
$^{4}$ Theoretical Physics Division, Institute of High Energy
Physics, Chinese Academy of Sciences, Beijing 10049, P.R. China\\
$^{5}$ Theoretical Physics Center for Science Facilities (TPCSF),
Chinese Academy of Sciences, P.R. China \\
$^{6}$ Kavli Institute for Theoretical Physics China, Chinese
Academy of Sciences, Beijing 100190, P.R. China\\
$^{7}$ Theory Division, CERN, CH-1211 Geneva, Switzerland \\
$^{8}$Department of Physics and Astronomy, University of Utah,
Salt Lake City, UT 84112, USA \\
$^{9}$ Key Laboratory of Frontiers in Theoretical Physics,
      Institute of Theoretical Physics, Chinese Academy of Sciences,
Beijing 100190, P. R. China \\
$^{10}$ George P. and Cynthia W. Mitchell Institute for
Fundamental Physics, Texas A$\&$M University, College Station, TX
77843, USA }

\begin{abstract}

In this paper we propose a dark matter model and study aspects of its
phenomenology. Our model is based on a new dark matter sector
with a $U(1)'$ gauge symmetry plus a discrete symmetry added to
the Standard Model of particle physics. The new fields of the dark
matter sector have no hadronic charges and couple only to leptons.
%The discrete symmetry ensures the existence of two degenerate
%stable dark matter particles.
Our model can not only give rise to the observed neutrino mass hierarchy,
but can also generate the baryon number asymmetry via non-thermal leptogenesis.
The breaking of the new $U(1)'$ symmetry produces cosmic
strings. The dark matter particles are produced non-thermally from
cosmic string loop decay which allows one to obtain sufficiently
large annihilation cross sections to explain the observed cosmic
ray positron and electron fluxes recently measured by the PAMELA, ATIC,
PPB-BETS, Fermi-LAT, and HESS
experiments while maintaining the required overall dark
matter energy density. The high velocity of the dark matter
particles from cosmic string loop decay leads to a low phase space
density and thus to a dark matter profile with a constant density
core in contrast to what happens in a scenario with thermally
produced cold dark matter where the density keeps rising towards
the center. As a result, the flux of $\gamma$ rays radiated from
the final leptonic states of dark matter annihilation from the
Galactic center is suppressed and satisfies the constraints from
the HESS $\gamma$-ray observations.

%which can simultaneously explain the cosmic ray anomalies recently
%measured by the PAMELA and ATIC experiments, yield the observed
%neutrino mass hierarchy, resolve the small-scale structure
%problems of the standard $\Lambda$CDM paradigm and satisfy
%constraint of $\gamma$ ray radiation by HESS observation.

\end{abstract}

\pacs{14.60.Pq, 95.35.+d}

\preprint{MIFP-09-21}

\maketitle

\section{Introduction}

There is strong evidence for the existence of a substantial amount of cold
dark matter (CDM). The leading CDM candidates are weakly
interacting massive particles (WIMPs), for example, the lightest neutralino
in supersymmetric models with $R$ parity. With a small
cosmological constant, the CDM scenario is consistent
with both the observations of the large scale structure of the Universe
(scales much larger than $1$Mpc) and the fluctuations of the
cosmic microwave background~\cite{BOPS}.

However, the collisionless CDM scenario  predicts too much power on small scales,
such as a large excess of dwarf galaxies~\cite{klypin,moore},
the over-concentration of dark matter (DM) in dwarf
galaxies~\cite{moore94,burkert,MB} and in large galaxies~\cite{NS}.
To solve this problem,  two of us with their collaborators
proposed  a scenario based on non-thermal production of WIMPs, which
 can be relativistic when generated. The WIMPs' comoving free-streaming
scales could be as large as or possibly even larger than 0.1 Mpc. Then,
the density fluctuations on scales less than the free-streaming scale would
be suppressed~\cite{zhang}. Thus,
the discrepancies between the observations of DM halos on
sub-galactic scales and the predictions of the standard WIMP DM
picture could be resolved.

Recently, the ATIC \cite{Chang:2008zz} and PPB-BETS
\cite{Torii:2008} collaborations have reported measurements of the
cosmic ray electron/positron spectrum at energies of up to $\sim 1$~TeV.
The data shows an obvious excess over
the expected background for energies in the ranges $\sim 300-800\,\textrm{GeV}$
and $\sim 500-800\,\textrm{GeV}$, respectively. At the same time, the PAMELA
collaboration also released their first cosmic-ray measurements of
the positron fraction \cite{Adriani:2008zr} and the $\bar{p}/p$
ratio \cite{Adriani:2008zq}. The positron fraction (but not the antiproton to proton ratio)
shows a significant excess for energies above $10\,\textrm{GeV}$ up to $\sim
100\,\textrm{GeV}$, compared to the background predicted by
conventional cosmic-ray propagation models. This result is
consistent with previous measurements by HEAT
\cite{Barwick:1997ig} and AMS \cite{Aguilar:2007yf}.

Very recently, the Fermi-LAT collaboration has released data on the
measurement of the electron spectrum from 20 GeV to 1 TeV
\cite{Fermi:2009zk}, and the HESS collaboration has published electron
spectrum data from 340 GeV to 700 GeV \cite{Aharonian:2009ah},
complementing their earlier measurements at 700 GeV to 5 TeV
\cite{HESS:2008aa}. The Fermi-LAT measured spectrum agrees with ATIC
below 300 GeV; however, it does not exhibit the special features at
large energy. There have already been some discussions on the implications
for DM physics obtained by combining the Fermi-LAT, HESS and
PAMELA results~\cite{fermidm}.

The ATIC, PPB-BETS and PAMELA results indicate the existence of a
new source of primary electrons and positrons, while the hadronic
processes are suppressed. It is well known that DM annihilation
can be a possible origin for primary cosmic rays \cite{signal}
which could account for the ATIC, PPB-BETS and PAMELA data
simultaneously, as discussed first in \cite{first} and also in
\cite{later} (see \cite{string} for a list of references)
\footnote{Note, however, that there are also astrophysical (see
e.g. \cite{astro}) or other particle physics (see e.g.
\cite{string}) explanations.}. However, the fact that the
$\bar{p}/p$ ratio does not show an excess gives strong constraints
on DM models if they are to explain the data. In particular, it is
very  difficult to use well-known DM candidates like the
neutralino to explain the ATIC and PAMELA data
simultaneously~\cite{donato} since they would also yield an excess
of antiprotons. Therefore, if the observed electron/positron or
positron excesses indeed arise from DM annihilation, it seems to
us that there may exist special connections between the DM sector
and lepton physics~\cite{Bi:2009md} (see also \cite{lepto1,lepto2}).

In this paper, we propose a DM model and study its implications
for DM detection. We fit our model to
two different combinations of the experiment data: one set of data
from the ATIC, PPB-BETS and PAMELA experiments; the other
from the Fermi-LAT, HESS, and PAMELA experiments.
Our results show that our model
can naturally explain the $e^\pm$ excesses while at the same time
solving the small scale problems of the standard $\Lambda$CDM
model via non-thermal DM production. For a single Majorana DM
particle, its annihilation cross section has $s$ wave
suppression. Thus, we consider two degenerate Majorana DM
particles. We add a new DM sector with a $U(1)'$ gauge
symmetry and introduce an additional discrete symmetry to the
Standard Model (SM). The DM particles are stable due to
the discrete symmetry. During the $U(1)'$ gauge symmetry breaking
phase transition a network of cosmic strings is generated. The
decay of cosmic string loops is a channel for producing a
non-thermal distribution of DM. This non-thermal
distribution allows for DM masses and annihilation cross
sections large enough to explain the cosmic ray anomalies while
simultaneously remaining consistent with the observed DM
energy density. In addition,  the observed neutrino masses and
mixings can be explained via the seesaw mechanism, and the baryon
number asymmetry can be generated via non-thermal
leptogenesis~\cite{Jeannerot:1996yi}.

It has been recently recognized  that a large annihilation cross
section of DM particles into leptons to account for the
cosmic ray anomalies will induce a large flux of $\gamma$ rays from the
Galactic Center (GC)~\cite{berg} or from the centers of dwarf
galaxies \cite{essig}. The predicted $\gamma$ ray fluxes based on
the NFW profile for the standard CDM scenario have been shown to be
in slight conflict with the current observations
of HESS \cite{hess}. However, in our model the DM particles are produced
non-thermally, so the high velocity of the DM particles will lower the
phase space density of DM and lead to a DM profile with a constant
density core \cite{kaplinghat}. Therefore our model with non-thermally
produced DM on one hand gives rise to a large annihilation cross section
to account for the positron/electron excess observed locally while on
the other hand it suppresses the DM density at the GC and leads to a low
flux of $\gamma$ ray radiation.

Our paper is organized as follows: in Section II, we describe
in detail the model and the production mechanism of the DM particles.
In Section III we study aspects of the
phenomenology of the model, including studies of some constraints
on the model parameters from particle physics experiments,
implications for the PAMELA, ATIC, PPB-BETS, Fermi-LAT, and HESS
results, and also the $\gamma$-ray radiation from the GC.
Section IV contains the discussion and conclusions.

\section{The Dark Matter Model}

\subsection{The Dark Matter Sector}

The DM model we propose consists of adding a new
``DM sector" to the Standard Model. The new particles
have only leptonic charges and are uncharged under color. This
ensures that the DM particles annihilate preferentially into
leptons. To ensure the existence of a stable DM particle, the new
sector is endowed with a discrete symmetry which plays a role
similar to that of R-parity in supersymmetric models. The lightest
particles which are odd under the $Z_2$ symmetry which we introduce
are the candidate DM particles.

In our convention, we denote the right-handed leptons
and Higgs doublet as $e^i_{R}(\textbf{1},-1)$ and
$H(\textbf{2},-\frac{1}{2})=\displaystyle{(H^{0}_{},H^{-}_{})^{T}_{}}$,
respectively, where their $SU(2)_L\times U(1)_Y$ quantum numbers are
given in parenthesis.

We consider the generalized Standard Model with an additional
$U(1)'$ gauge symmetry broken at an intermediate scale. In particular,
all the SM fermions and Higgs fields
are uncharged under this $U(1)'$ gauge symmetry.
To break the $U(1)'$ gauge symmetry, we introduce a SM singlet Higgs
field $S$ with $U(1)'$ charge $\mathbf{-2}$. Moreover, we introduce
four SM singlet chiral fermions $\chi_1$, $\chi_2$, $N_1$, and $N_2$,
a SM singlet scalar field $\widetilde{E}$ and a SM doublet
scalar field $H'$ with $SU(2)_L\times U(1)_Y$
quantum numbers $(\mathbf{1}, \mathbf{-1})$ and $(\mathbf{2},
\mathbf{\frac{1}{2}})$, respectively.
The $U(1)'$ charges for $\chi_i$ and $H'$ are
$\mathbf{1}$, while the $U(1)'$ charges for $N_i$ and $\widetilde{E}$
are $\mathbf{-1}$. Thus, our model is anomaly free.
To have stable DM candidates, we introduce a ${\bf Z}_2$ symmetry.
Under this ${\bf Z}_2$ symmetry, only the particles
$\chi_i$ and $\widetilde{E}$ are odd while all the other particles are even.
The $\chi$ particles will be the DM candidates, whereas the
chiral fermions $N_i$ will play the role of right-handed neutrinos.

The relevant part of the most general renormalizable Lagrangian consistent
with the new symmetries is
\begin{eqnarray}
 -{\cal L}& =& {1\over 2} m_S^2 S^{\dagger} S +
{1\over 2} m^2_{\widetilde{E}} \widetilde{E}^{\dagger} \widetilde{E}
+ {1\over 2} m_{H'}^2 H^{\prime \dagger} H'
+ {{\lambda}\over 4} (S^{\dagger} S)^2
+ {{\lambda_1}\over 4}
(\widetilde{E}^{\dagger} \widetilde{E})^2  + {{\lambda_2}\over 4}
(H^{\prime \dagger} H')^2
\nonumber\\ &&
 + {{\lambda_3}\over 2}
(S^{\dagger} S) (\widetilde{E}^{\dagger} \widetilde{E})
 + {{\lambda_4}\over 2}  (\widetilde{E}^{\dagger}
\widetilde{E})
 (H^{\prime \dagger} H')
+{{\lambda_5}\over 2} (S^{\dagger} S) (H^{\prime \dagger} H')
+ {{\lambda_6}\over 2} (S^{\dagger} S) (H^{\dagger} H)
\nonumber\\ &&
+ {{\lambda_7}\over 2} (\widetilde{E}^{\dagger}
\widetilde{E}) (H^{\dagger} H)
+ {{\lambda_8}\over 2} (H^{\prime \dagger} H')(H^{\dagger} H)
 + \left( y_e^{i}\overline{ e_{R}^{i} }
\widetilde{E} \chi_1 +  y_e^{\prime i}\overline{ e_{R}^{i} }
\widetilde{E} \chi_2
\right.\nonumber\\ && \left.
+ y_{\chi}^{ij} S \overline{\chi^c_i} \chi_j
 + y_{N}^{ij} S^{\dagger} N_i N_j
+ y_{\nu}^{ij} L_i H' N_j
 +\textrm{H.c.} \right)~.~
\label{DAMA-I}
\end{eqnarray}

As we will discuss in the following subsection, the vacuum expectation
value (VEV) for $S$ is around $10^{9}$ GeV. Then, the couplings
$\lambda_3$, $\lambda_5$ and $\lambda_6$ should be very small - about
$10^{-12}$ - in order for the model to be consistent with the expected
value of the SM Higgs. This fine-tuning problem could be solved naturally
if we were to consider a supersymmetric model.
Moreover, in order to explain the recent cosmic ray data,
the Yukawa couplings $y_{\chi}^{ij}$ should be around
$10^{-6}$. This would generate a DM mass around 1 TeV.
Such small Yukawa couplings $y_{\chi}^{ij}$ can be explained via
the Froggat-Nielsen mechanism~\cite{Froggatt:1978nt}
which will not be studied here.

To explain the neutrino masses and mixings via the ``seesaw
mechanism", we require that the VEV of $H'$ be about 0.1 GeV
if $y_N^{ij} \sim 1$ and $y_{\nu}^{ij} \sim 1$.
In this case, the lightest active
neutrino is massless since we only have two right-handed neutrinos
$N_i$. In addition,  in our $U(1)'$ model,
 the Higgs field forming the strings is also the
Higgs field which gives masses to the right-handed neutrinos.
There are right-handed neutrinos trapped as transverse zero modes
in the core of the strings.
When cosmic string loops decay, they release these neutrinos. This is an
out-of-equilibrium process. The released neutrinos acquire heavy
Majorana masses
and decay into massless leptons and electroweak Higgs particles to produce a
lepton asymmetry, which is converted into a baryon number
asymmetry via sphaleron transitions~\cite{Jeannerot:1996yi}.
Thus, we can explain the baryon number asymmetry
via non-thermal leptogenesis.

In this paper, we consider two degenerate
Majorana DM candidates $\chi_1$ and $\chi_2$ since
the annihilation cross section for a single Majorana DM particle is too
small to explain the recent cosmic ray experiments~\cite{Bi:2009md}.
For simplicity, we assume that  the Lagrangian is invariant under
$\chi_1 \leftrightarrow \chi_2$. Thus, we have
\begin{eqnarray}
y_e^{i} \equiv y_e^{\prime i}~,~~~y_{\chi}^{ij} \equiv y_{\chi}^{ji} ~.~\,
\end{eqnarray}
To make sure that we have  two degenerate
Majorana DM candidates $\chi_1$ and $\chi_2$,
we choose $y_{\chi}^{12}=0$, and assume $m_{\chi} < m_{\widetilde{E}}$.

\subsection{Non-Thermal Dark Matter Production via Cosmic Strings}

We assume that the $U(1)'$ gauge symmetry is broken by the
VEV of the scalar field $S$. To be specific, we take the potential
of $S$ to be
\be
V(S) \, = \, \frac{1}{4} \lambda \bigl( |S|^2 - \eta^2 \bigr)^2 \, ,
\ee
where $\lambda$ is the self-interaction coupling constant. The VEV
of $S$ hence is $\langle S \rangle = \eta$ with $m^2_S=\lambda
\eta^2$. Due to finite temperature effects, the symmetry is
unbroken at high temperatures. During the cooling of the very
early universe, a symmetry breaking phase transition takes place
at a temperature $T_c$ with
\be
T_c \, \simeq \, \sqrt{\lambda} \eta \, .
\ee
During this phase transition, inevitably a network of local cosmic strings
will be formed. These strings are topologically non-trivial field configurations
formed by the Higgs field $S$ and the $U(1)'$ gauge field $A$.
The mass per unit length of the strings is given by $\mu = \eta^2$.

During the phase transition, a network of strings forms, consisting of both
infinite strings and cosmic string loops. After the transition, the infinite
string network coarsens and more loops form from the
intercommuting of infinite strings. Cosmic string loops loose their energy by
emitting gravitational
radiation. When the radius of a loop becomes of the order of the string
width $w \simeq \lambda^{-1/2} \eta^{-1}$,
the loop releases its final energy into a burst of $A$ and $S$
particles \footnote{We are not considering here DM production from
cosmic string cusp annihilation since the efficiency of this mechanism
may be much smaller than the upper estimate established in
\cite{RHBcusp}, as discussed e.g. in \cite {Olum}. DM production
from cusp annihilation has been considered in \cite{Morissey}.}.
Those particles subsequently decay into DM particles,  with
branching ratios $\epsilon$ and $\epsilon'$. For simplicity we
assume that all the final string energy goes into $A$ particles. A single
decaying cosmic string loop thus releases
\be
N  \, \simeq \, 2 \pi \lambda^{-1} \epsilon
\ee
DM particles which we take to have a monochromatic distribution with energy
$E \sim {T^c \over 2}$, the energy of an S-quantum in the broken phase.
In our model, we assume that the masses for $A$,
$S$ and $N_i$ are roughly the same, so we have $\epsilon=1$.

Given the symmetry we have imposed, the number densities of
$\chi_1$ and $\chi_2$ are equal. Thus, the number density
$n_{DM}$ of DM particles, the sum of the number densities of $\chi_1$ and
$\chi_2$, is
\begin{eqnarray}
n_{DM} \, \equiv \, n_{\chi_1} + n_{\chi_2} \, = \, 2n_{\chi_1} \, = \, 2n_{\chi_2} ~.~\,
\end{eqnarray}

If the $S$ and $A$ quanta were in thermal equilibrium before the phase
transition, then the string network is formed with a microscopic
correlation length $\xi(t_c)$ (where $t_c$ is the time at which the
phase transition takes place). The correlation length gives the mean
curvature radius and mean separation of the strings. As discussed
in \cite{Kibble} (see also the reviews \cite{ShelVil}), the initial
correlation length is
\be
\xi(t_c) \, \sim \, \lambda^{-1} \eta^{-1} \, .
\ee

After string formation, there is a time interval during which the dynamics
of the strings is friction-dominated. In this period, the correlation length
increases faster than the Hubble radius because loop intercommutation
is very efficient. As was discussed e.g. in \cite{RobRiotto}, the correlation
length scale $\xi(t)$ in the friction epoch scales as
\begin{equation} \label{friction}
\xi(t) \, = \, \xi(t_c) \left ({t \over t_c}\right )^{3\over 2}
~.~\,
\end{equation}
The friction epoch continues until $\xi(t)$ becomes comparable to the
Hubble radius $t$. After this point, the string network follows a
``scaling solution" with $\xi(t) \sim t$. This scaling solution continues
to the present time.

The loss of energy from the network of long strings with correlation length
$\xi(t)$ is predominantly due to the production of cosmic string loops.
The number density of cosmic string loops
created per unit of time is
given by \cite{ShelVil,RobRiotto}:
\begin{equation} \label{prod}
{dn\over dt} \, = \, \nu \xi^{-4} {d\xi \over dt} ~,~\,
\end{equation}
where $\nu$ is a constant of order 1. We are interested in loops
decaying below the temperature $T_\chi$ when the DM
particles fall out of thermal equilibrium (loops decaying earlier
will produce DM particles which simply thermalize). We
denote the corresponding time by $t_{\chi}$.

The DM number density released from $t_{\chi}$ till today is
obtained by \cite{zhang} summing up the contributions of all
decaying loops. Each loop yields a number $N$ of DM
particles. We track the loops decaying at some time $t$ in terms
of the time $t_f$ when that loop was created. Since the loop
density decreases sharply as a function of time, it is the loops
which decay right after $t_{\chi}$ which dominate the
integral. For the values of $G \mu$ which we are interested
in, it turns out that loops decaying around $t_{\chi}$ were created
in the friction epoch, and the loop number density is determined by
inserting (\ref{friction}) into (\ref{prod}). Changing the integration
variable from $t$ to $\xi(t)$, we
integrate the redshifted number
density to obtain:
\begin{equation}
n^{nonth}_{DM}(t_0) \, = \, N \nu \int^{\xi_0}_{\xi_F} \left ( {t \over t_0} \right
)^{3\over 2} \xi^{-4} d\xi ~,~\, \label{eq:no}
\end{equation}
where the subscript $0$ refers to parameters which are
evaluated today. In the above,
$\xi_F = \xi(t_F)$ where $t_F$ is the time at which cosmic string
loops which are decaying at the time $t_{\chi}$ formed.

Now the loop's time-averaged radius (radius averaged over a
period of oscillation) shrinks at a rate \cite{ShelVil}
\be
{dR\over dt} \, = \, - \Gamma_{loops} G \mu \, ,
\ee
where $\Gamma_{loops}$ is a numerical factor $\sim 10-20$. Since loops
form at time $t_F$ with an average radius
\be
R(t_F) \, \simeq \, \lambda^{1/2} {g^{*}}^{3/4} G \mu M_{pl}^{1\over 2} t_F^{3\over 2},
\ee
where $g_*$ counts the number of massless degrees of
freedom in the corresponding phase,
they have shrunk to a point at the time
\be
t  \, \simeq \, \lambda^{1/2}  {g^{*}}^{3/4} \Gamma^{-1}_{loops} M_{\rm Pl}^{1\over 2} t_F^{3\over 2}.
\ee
Thus
\be
t_F \, \sim \, \lambda^{-1/3}  {g^{*}}^{-1/2} \Gamma^{2\over 3}_{loops} M_{\rm Pl}^{-{1\over 3}}
t_{\chi}^{2\over 3}.
\ee

Now the entropy density is
\be
s \, = \, {2 \pi^2 \over 45} g_* T^3 \, .
\ee
The time $t$ and temperature $T$ are related by
\be
t \, = \, 0.3 g_*^{-{1\over 2}}(T) {M_{\rm Pl}\over T^2} \, ,
\ee
where $M_{\rm Pl}$ is the Planck
mass. Thus using Eqs.~(\ref{friction}) and (\ref{eq:no}), we find
that the DM number density today released by decaying cosmic string
loops is given by
\begin{equation}
Y^{nonth}_{DM} \equiv \, {n^{nonth}_{DM}\over s} \, = \, {{6.75} \over {\pi}} \epsilon \nu
\lambda^{3/2} \Gamma_{loops}^{-2} g_{*_{T_c}}^{3/2}
g_{*_{T_\chi}} g_{*_{T_{F}}}^{-5/2}
 M_{\rm Pl}^2\, {T_{\chi}^4 \over T_c^6} \, , \label{eq:Ynonth}
\end{equation}
where the subscript on $g^*$ refers to the time when $g^*$ is evaluated.

The DM relic abundance is related to $Y_{\chi}$ by:
\begin{eqnarray}
\Omega_\chi \, h^2 & \approx & m_{\chi} Y_{\chi} s(t_0) \rho_c(t_0)^{-1} h^2 \nonumber \\
& \approx & 2.82 \times 10^8\, Y^{tot}_\chi\,
(m_{\chi}/{\rm GeV}) ~,~\, \label{eq:Omega}
\end{eqnarray}
where $h$ is the Hubble parameter in units of $100 {\rm km}{\rm s}^{-1} {\rm Mpc}^{-1}$,
$m_\chi$ is the DM mass,
and $Y^{tot}_{\chi} =Y^{therm}_\chi+ Y^{nonth}_\chi$.

To give some concrete numbers, we choose the
parameter values $\epsilon=1$, $\nu=1$, $\lambda=0.5$,
$\Gamma=10$,  $M_{\rm Pl}=1.22\times 10^{19}~{\rm GeV}$
and $\Omega_\chi\, h^2 =0.11$. In our model,
we have $g_{*_{T_c}}=136$, $g_{*_{T_F}}=128$,
 and $g_{*_{T_\chi}}=128$.
We define the dimensionless ratios
\begin{eqnarray}
\alpha \equiv {{m_{\chi}}\over {T_{\chi}}}~,~~~
\beta \equiv {{Y^{nonth}_\chi}\over {Y^{tot}_{\chi}}}~.~\,
\end{eqnarray}
Demanding that we obtain a specific value of $\beta$ for the above choices of the
parameter values will fix $T_c$ via (\ref{eq:Omega}).
For various values of $\alpha$ and $\beta$, we present the resulting $T_c$ values for the cases $m_{\chi}=620~{\rm GeV}$,
$m_{\chi}=780~{\rm GeV}$,
and $m_{\chi}=1500~{\rm GeV}$, respectively, in Table~\ref{Model-I-II}.
In short, $T_c$ must be around $10^9$ GeV if we want to
generate enough DM density non-thermally via cosmic strings.

\begin{table}[t]
\caption{The required  $T_c$ values
in units of GeV for various choices of
$\alpha$ and $\beta$ in the cases
$m_{\chi}=620~{\rm GeV}$, $m_{\chi}=780~{\rm GeV}$,
and $m_{\chi}=1500~{\rm GeV}$, respectively. }
\renewcommand{\arraystretch}{1.25}
\begin{center}
\begin{tabular}{|c|c|c|c|c|c|c|}
\hline
$\alpha$ & 1 & 1 & 2 & 2 & 5 & 5  \\ \hline
$\beta $ & 1 & 0.5 & 1 & 0.5 & 1 & 0.5 \\ \hline
$T_c$ ($m_{\chi}=620~{\rm GeV}$) & $7.7\times 10^9$ & $8.6\times 10^9$ &
$4.8\times 10^9$ & $5.4\times 10^9$ &
$2.6\times 10^9$ & $2.9\times 10^9$  \\ \hline
$T_c$ ($m_{\chi}=780~{\rm GeV}$) & $9.3\times 10^9$ & $1.0\times 10^{10}$ &
$5.9\times 10^9$ & $6.6\times 10^9$ &
$3.2\times 10^9$ & $3.6\times 10^9$  \\ \hline
$T_c$ ($m_{\chi}=1500~{\rm GeV}$) & $1.6\times 10^{10}$ & $1.8\times 10^{10}$ &
$1.0\times 10^{10}$ & $1.1\times 10^{10}$ &
$5.5\times 10^9$ & $6.2\times 10^9$  \\
\hline \hline
$\alpha$ & 10 & 10 & 15 & 15 & 20 & 20  \\ \hline
$\beta $ & 1 & 0.5 & 1 & 0.5 & 1 & 0.5 \\ \hline
$T_c$ ($m_{\chi}=620~{\rm GeV}$) & $1.7\times 10^9$ & $1.9\times 10^9$ &
$1.3\times 10^9$ & $1.4\times 10^9$ &
$1.0\times 10^9$ & $1.2\times 10^9$ \\ \hline
$T_c$ ($m_{\chi}=780~{\rm GeV}$) & $2.0\times 10^9$ & $2.2\times 10^9$ &
$1.5\times 10^9$ & $1.7\times 10^9$ &
$1.3\times 10^9$ & $1.4\times 10^9$ \\ \hline
$T_c$ ($m_{\chi}=1500~{\rm GeV}$) & $3.5\times 10^9$ & $3.9\times 10^9$ &
$2.6\times 10^9$ & $3.0\times 10^9$ &
$2.2\times 10^9$ & $2.4\times 10^9$ \\ \hline
\end{tabular}
\end{center}
\label{Model-I-II}
\end{table}

\section{Phenomenology of the model}
%\subsection{ Explanations for the ATIC and PAMELA Results}
\subsection{ Constraints on the Model Parameters}

%1, free-streaming

%2, constraint from g-2 on e, mu\\
The coupling constants $y_e^i$ between right-handed leptons and
the DM sector are constrained by experiments, and
especially by the precise value of muon anomalous magnetic moment $g-2$. Assuming that the
masses of $\chi$ and $\tilde{E}$ are nearly degenerate, we obtain
that the contribution to the muon anomalous magnetic moment from
the new coupling is about \cite{bi02}
\be
\delta a_i  \, \sim \, (y_e^i)^2 \frac{1}{192 \pi^2}
\frac{m_{e^i}^2}{m_\chi^2}  ~.~\,
\ee
The $2\sigma$ upper bound from the E821 Collaboration on $\delta
a_\mu$ is smaller than $\sim 40\times 10^{-10}$~\cite{e821}, from
which we get for $m_\chi \sim 1$ TeV,
\be
y_\mu \, \lesssim \, 10 \, .
\ee
For the electron anomalous magnetic momentum we assume the
contribution from the dark sector is within the experimental
error~\cite{pdg}
\be
\delta a_e \, \le \, 7 \times 10^{-13} \,  .
\ee
Then we get a upper limit on $y_e$ which is about $30$. Therefore
the constraints on the couplings of the model due to the heavy
masses of the new particles are quite loose.

%3, LFV

 Now we study the constraints from the experimental limits on
 lepton flavor violation (LFV) processes
%rom the Lagrangian in Eq. (1) we notice that the dark sector may
%induce lepton flavor (LFV) violation processes,
such as $\mu\to e
\gamma$, $\tau\to\mu(e)\gamma$ and so on. The branching ratios for
the radiative LFV processes are given by \cite{bi02}
\be
Br(e_i\to e_j \gamma) \, \sim \, \alpha_{em} m_i^5/2 \times \left(
\frac{Y_e^iy_e^j}{384\pi^2m_\chi^2}\right)^2/\Gamma_i \, ,
\ee
where $\Gamma_i$ is the width of $e_i$. Given the experimental
constraint on the process $\mu\to e \gamma$ we get
\be
Br(\mu\to e \gamma) \, \sim \, 10^{-8}\times (y_ey_\mu)^2 \lesssim 10^{-11}~,~\,
\ee
which gives that $y_ey_\mu \lesssim 0.03$. For the process $\tau \to
\mu(e) \gamma$ we have
\be
Br(\tau\to \mu(e) \gamma) \, \sim \, 10^{-9}\times
(y_{\mu (e)} y_\tau)^2 \lesssim 10^{-7}~,~\,
\ee
which leads to the conclusion that $y_\tau y_{\mu (e)} \lesssim
10$. Connecting the DM sector to the PAMELA and Fermi-LAT (or
ATIC) results usually requires a large branching ratio into
electron and positron pairs. From the LFV constraints shown above
we conclude that it is possible to have a large branching ratio
for the annihilation of the DM particles directly into
$e^+ e^-$, or via $\mu^+ \mu^-$.

\subsection{Explanation for the Cosmic $e^\pm$ Excesses}

In our model the DM sector only couples to the SM
lepton sector. Therefore DM annihilates into leptons dominantly.
Furthermore, since DM is produced non-thermally in our model the
DM annihilation rates can be quite large with a sizable Yukawa
coupling $y_e^i$. Thus our model can naturally explain the cosmic
$e^\pm$ excesses observed.

 Because the annihilation cross sections for $\chi_1 \chi_1$ and
$\chi_2 \chi_2$ to leptons are $s$ wave suppressed, the dominant
cross sections of $\chi_1 \chi_2$ annihilating into charged
leptons are given by~\cite{Bi:2009md}
\begin{eqnarray}
\label{cross-section-majorana-1}
\sigma_{ij}^{} v &\equiv&
\sigma_{\chi_1\chi_2\rightarrow
e_R^i e_R^{cj}} v
\nonumber\\
&=& \frac{4}{32\pi}|y_{e}^{i}|^{2}_{}|y_{e}^{j}|^{2}_{}
\frac{1}{s\sqrt{s\left(s-4m_{\chi}^{2}\right)}}
\left\{\sqrt{s\left(s-4m_{\chi}^{2}\right)}+
\left[2\left(m_{\widetilde{E}}^{2}-m_{\chi}^{2}\right)
-\frac{2m_{\chi}^{2}s}{s+2m_{\widetilde{E}}^{2}-2m_{\chi}^{2}}\right]
\right.\nonumber\\&& \times
\ln\left|\frac{s+2m_{\widetilde{E}}^{2}
-2m_{\chi}^{2}-\sqrt{s\left(s-4m_{\chi}^{2}\right)}}
{s+2m_{\widetilde{E}}^{2}-2m_{\chi}^{2}
+\sqrt{s\left(s-4m_{\chi}^{2}\right)}}\right|
+2\left(m_{\widetilde{E}}^{2}-m_{\chi}^{2}\right)^{2}_{}
\nonumber\\&& \left.\times
\left[\frac{1}{s+2m_{\widetilde{E}}^{2}-2m_{\chi}^{2}
-\sqrt{s\left(s-4m_{\chi}^{2}\right)}}
-\frac{1}{s+2m_{\widetilde{E}}^{2}-2m_{\chi}^{2}
+\sqrt{s\left(s-4m_{\chi}^{2}\right)}}\right]\right\}~,~\,
\end{eqnarray}
where $v$ is the relative velocity between the two annihilating particles
in their center of mass system. The overall factor 4 will be cancelled when
we calculate the lepton fluxes, so, we will leave it
in our discussions.
 Up to $\mathcal{O}(v^{2}_{})$, the above cross
section  can be simplified as~\cite{Bi:2009md}
\begin{eqnarray}
\label{cross-section-majorana-2} \sigma_{ij}^{}
v&\simeq&\frac{4}{128\pi}|y_{e}^{i}|^{2}_{}|y_{e}^{j}|^{2}_{}
\left\{\frac{8}{(2+r)^{2}_{}}+\left[\frac{1}{(2+r)^{2}_{}}-\frac{8}{(2+r)^{3}_{}}\right]v^{2}_{}\right\}\frac{1}{m_{\chi}^{2}} ~,~\,
\end{eqnarray}
where
\begin{eqnarray}
r\equiv \frac{m_{\widetilde{E}}^{2}-m_{\chi}^{2}}{m_{\chi}^{2}}>0\,.
\end{eqnarray}
With $v \sim 10^{-3}_{}$ and $r\sim 0$, we obtain~\cite{Bi:2009md}
\begin{eqnarray}
\label{cross-section-dirac-3} \langle\sigma_{ij}^{}
v\rangle&\lesssim& 4\times 1.2\times
10^{-25}_{}\,\textrm{cm}^{3}_{}\textrm{sec}^{-1}_{}
\left(\frac{700\,\textrm{GeV}}{m_{\chi}^{}}\right)^{2}_{}
|y_{e}^{ i}|^{2}_{}|y_{e}^{ j}|^{2}_{}~.~\,
\end{eqnarray}
We emphasize that the Yukawa couplings $y_{e}^{i}$
 should be smaller than ${\sqrt {4\pi}}$
for the perturbative analysis to be valid.

%In order to explain the ATIC/PPB-BETS and PAMELA/HEAT data, we
%assume that the DM is produced by some non-thermal mechanism like
%the one reviewed in the previous subsection~\cite{zhang}.
In our model with non-thermal production of DM particles,
we consider two separate fits to the ATIC/PPB-BETS/PAMELA and
Fermi-LAT/HESS/PAMELA datasets.
Firstly we consider a numerical fit to the ATIC, PPB-BETS and PAMELA
 data~\cite{Bi:2009md}. In this case we assume the DM mass
to be $620$ GeV and that DM annihilates into electron/positron
pairs predominantly, {\it i.e.}, $y_e^{i}\sim 0$ for $i=2,~3$. In
the second case we fit the Fermi-LAT, HESS and PAMELA data
by taking the DM mass $1500$ GeV and assuming
%that DM annihilates into $e^+e^-$, $\mu^+\mu^-$ and $\tau^+\tau^-$
%with
% equal branching ratios.
that DM annihilates into $\mu^+\mu^-$ pairs dominantly. Note that
all lepton fluxes resulting from DM annihilation are proportional
to $n_{\chi}^2 \sigma_{\rm ann}$ for models with a single DM
candidate $\chi$. Because
$n_{\chi_1}=n_{\chi_2}=n_{\chi}/2$ in our model,  the lepton
fluxes are proportional to
\begin{eqnarray}
n_{\chi_1} n_{\chi_2} \sigma_{\rm ann} ~=~
{1\over 4} n_{\chi}^2 \sigma_{\rm ann}  ~.~\,
\end{eqnarray}
This will cancel the overall factor 4 in the above annihilation cross sections
in Eqs. (\ref{cross-section-majorana-1})
and (\ref{cross-section-majorana-2}).

\begin{figure}[!htb]
\begin{center}
\includegraphics[width=0.45\columnwidth]{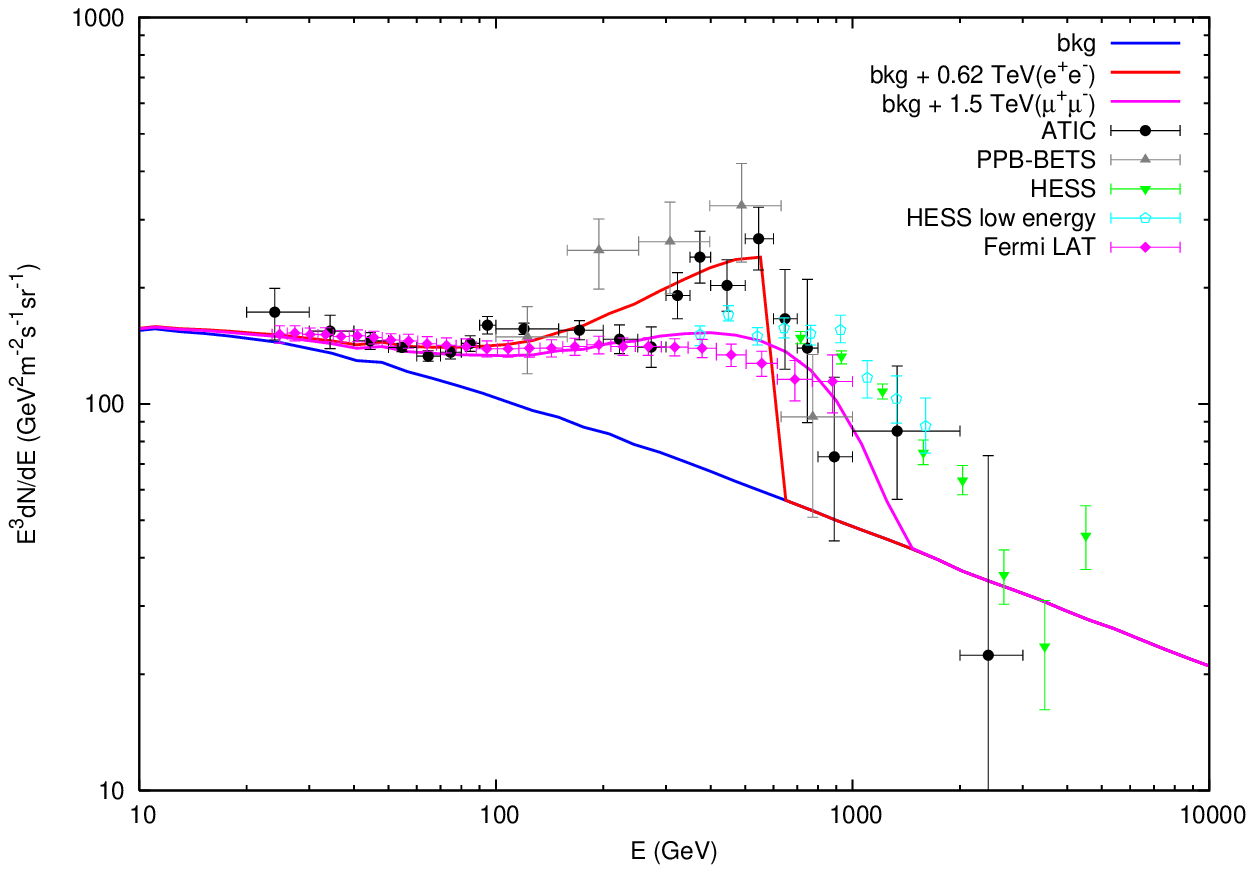}
\includegraphics[width=0.45\columnwidth]{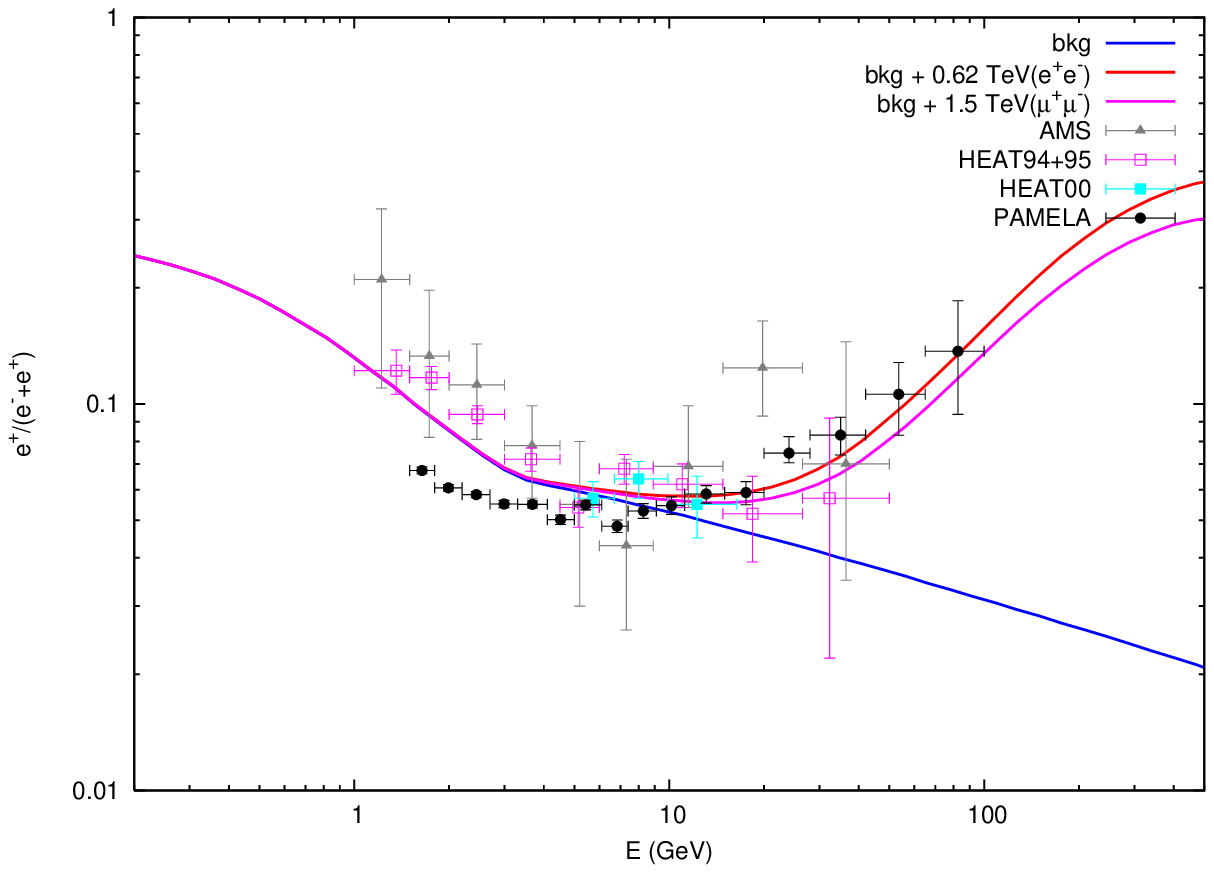}
\caption{Left: The $e^++e^-$ spectrum including the contribution from
DM annihilation compared with the observational data from ATIC
\cite{Chang:2008zz}, PPB-BETS \cite{Torii:2008}, HESS
\cite{HESS:2008aa,Aharonian:2009ah} and Fermi-LAT
\cite{Fermi:2009zk}. Right: The $e^+/(e^-+e^+)$ ratio including the
contribution from DM annihilation as a function of energy compared with
the data from AMS \cite{Aguilar:2007yf}, HEAT \cite{Barwick:1997ig,
Coutu:2001jy} and PAMELA \cite{Adriani:2008zr}. Two sets of fitting
parameters are considered: in one model (Model I) the DM mass is $620$ GeV
with $e^+e^-$ being the main annihilation channel to fit the ATIC data,
while in the other model (Model II) the DM mass is $1500$ GeV and we
assume that $\mu^+\mu^-$ is the main annihilation channel to fit the Fermi-LAT
data.}
\label{data}
\end{center}
\end{figure}

In Fig. \ref{data} we show that both cases can give a good fit to
the data after considering the propagation of electrons and
positrons in interstellar space~\cite{Bi:2009md} with the
annihilation cross section $0.75\times 10^{-23}\, \textrm{cm}^3
\textrm{s}^{-1}$ and $3.6\times 10^{-23} \,\textrm{cm}^3
\textrm{s}^{-1}$, respectively. The model parameters of the two
fits are given in Table \ref{table:cross}. For the first fit,
we do not need the boost factor at all by choosing $y_e^{1}=2.6$,
which is still smaller than the upper limit ${\sqrt
{4\pi}}$ for a valid perturbative theory.
Moreover,  choosing $y_e^{2}=3$ in the second
fit, we just need a small boost factor about 10
which  may be due to the clumps of
the DM distribution~\cite{yuan}.
% and boost
%factors ($BF$) of $26$ and $392$ are needed. However if we take
%$y_e^{1}=2.26$, which is still smaller than the limit ${\sqrt
%{4\pi}}$ for the applicability of perturbation theory, we do not
%need a non-trivial BF at all. Moreover, in the second case,
%according to Eq.~(\ref{cross-section-dirac-3}), we obtain that for
%order one (${\cal O}(1)$)  Yukawa couplings ( $y_\mu$ $\sim 2.5$),
%the boost factor can be as small as about $10$, which may be due
%to the clumping of the DM distribution~\cite{yuan}.
Therefore, the results on the observed cosmic $e^\pm$
excesses can be explained
naturally in our model.
% if the Yukawa couplings
%$y_e^{i}$ are around order one (${\cal O}(1)$).

\subsection{ $\gamma$-Ray Radiation from the Galactic Center }

Since the explanations of the anomalous cosmic ray require a very
large annihilation cross section to account for the
observational results, this condition leads to a strong
$\gamma$-ray radiation from the final lepton states. In
particular, observations of the GC \cite{berg}
or the center of dwarf galaxies \cite{essig} have already led to
constraints on the flux of the $\gamma$-ray radiation.

The HESS observation of $\gamma$-rays from the GC
\cite{hess} sets constraints on the Galactic DM profile. The
NFW profile in the standard CDM scenario leads to too large a flux of
$\gamma$-rays, thus conflicting with the HESS observation. On the
other hand, if DM is produced non-thermally as suggested in
Section II the DM profile will have a constant
density core \cite{kaplinghat} so that the $\gamma$-ray radiation from
the GC will be greatly suppressed.

In our numerical studies, we consider the following two
cases to constrain the DM profile:

\begin{itemize}

\item Case I: we simply require that the $\gamma$-ray flux
due to final state radiation (FSR) do not exceed the HESS
observation.

\item Case II: we make a global fit to the HESS data by
assuming an astrophysical source with power law spectrum plus an
additional component from FSR resulting from DM annihilation.

\end{itemize}

%Since the annihilation cross sections for the two seperate fits
%has been fixed to account for the experimental results, the FSR
%depends only on the DM profile. Therefore the HESS data will
%directly constrain the Galactic profile.

\begin{figure}[!htb]
\begin{center}
\includegraphics[width=0.45\columnwidth]{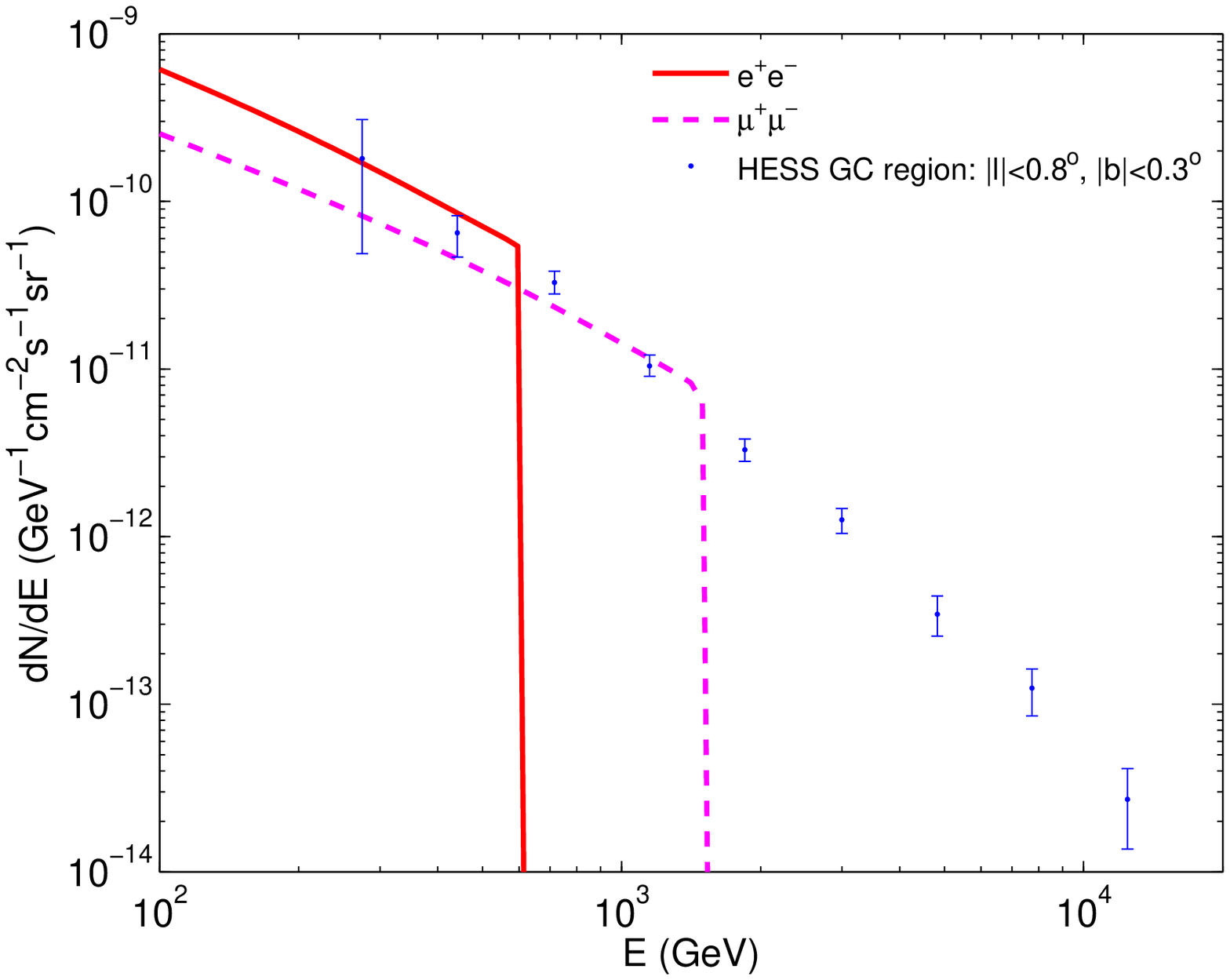}
\includegraphics[width=0.45\columnwidth]{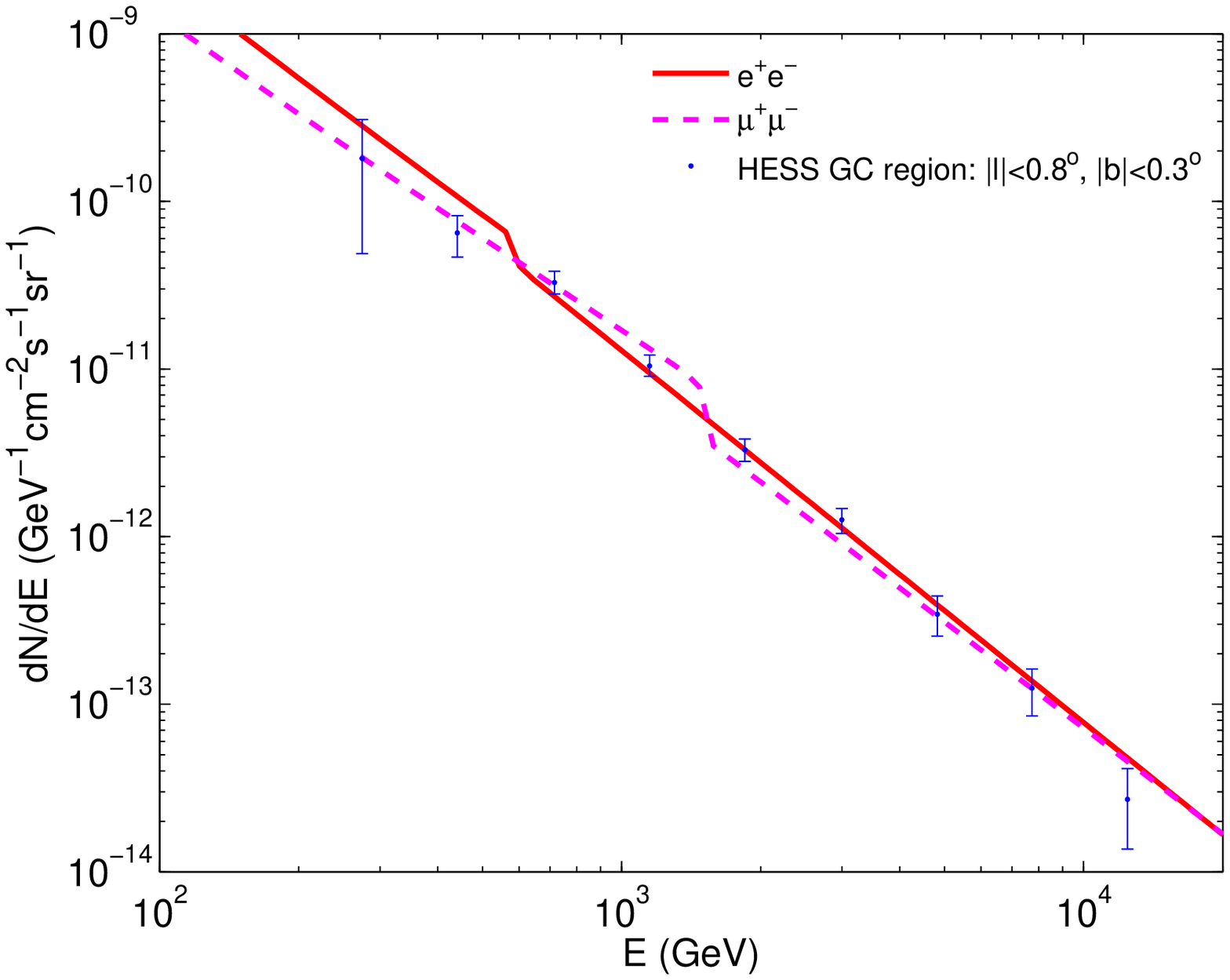}
\includegraphics[width=0.45\columnwidth]{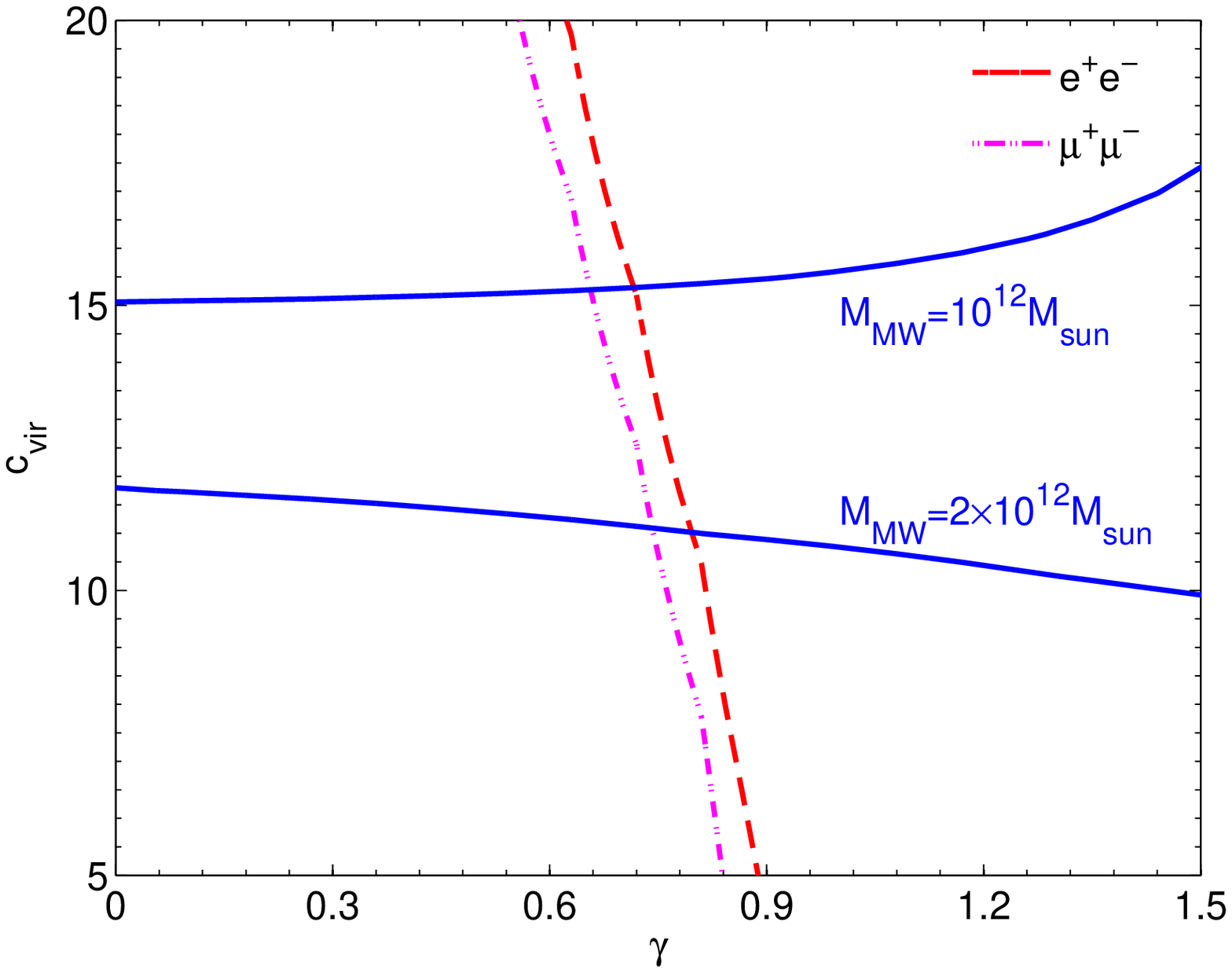}
\includegraphics[width=0.45\columnwidth]{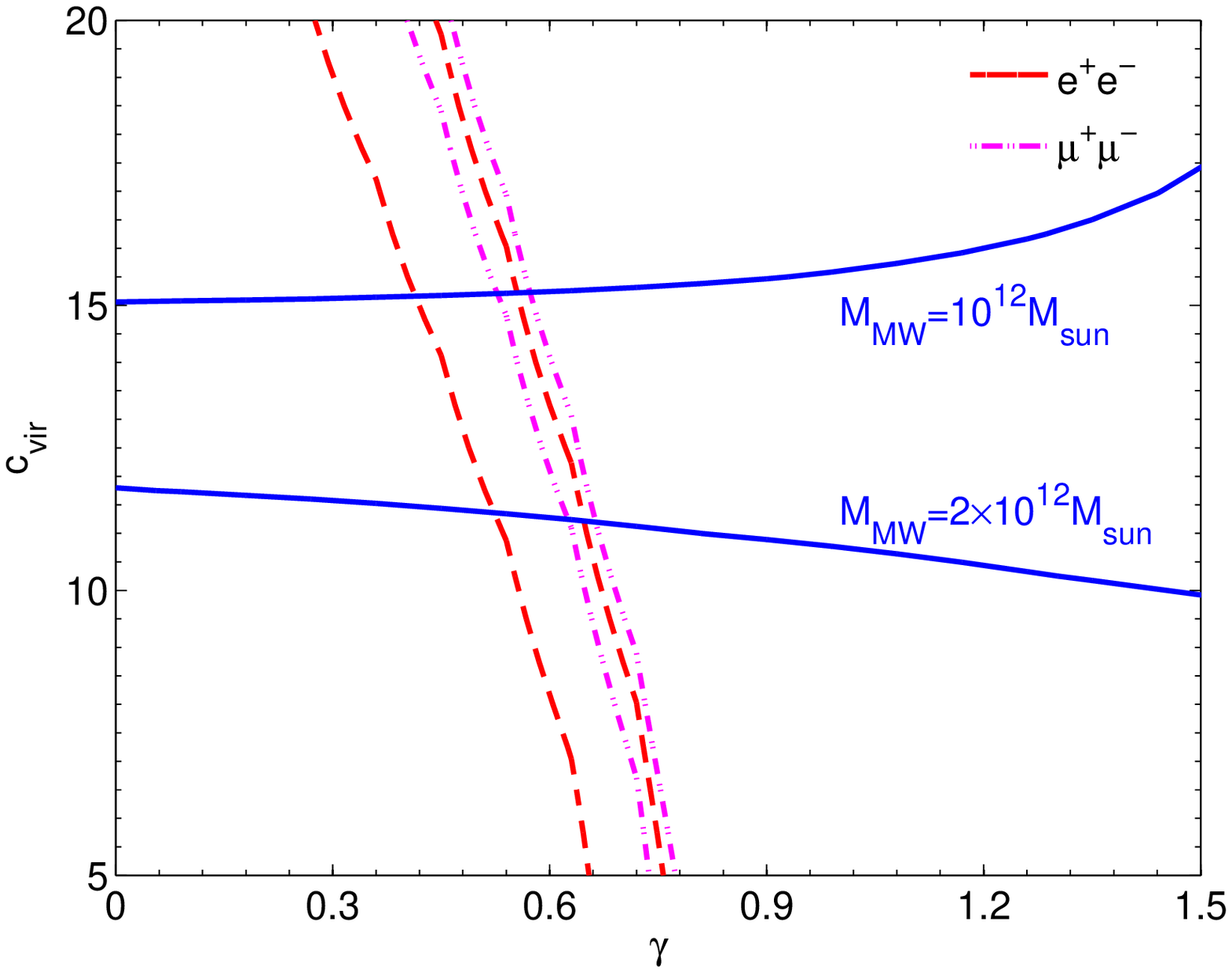}
\caption{Upper: the FSR $\gamma$-ray fluxes from a region with
$|l|<0.8^\circ$ and $|b|<0.3^\circ$ close to GC compared with the
observational data from HESS \cite{hess}. The left panel compares
the two models given in Table \ref{table:cross} directly with the data, while the right panel shows the combined fitting results
using a power law astrophysical background together with the FSR
contribution from DM annihilation at $95\%$ ($2\sigma$) confidence
level.
Lower: constraints on the DM profile parameters $\gamma$ and
$c_{vir}$ due to the HESS observation of $\gamma$-ray radiation
from the GC by assuming different final leptonic states. The left
panel corresponds to the constraint Case I, while the right panel
corresponds to Case II. The two curves in the right panel represent
the $1\sigma$ and $2\sigma$ upper bounds respectively.
}
\label{gc}
\end{center}
\end{figure}

Let us consider a DM profile taking the form
\begin{equation}
\label{pro} \rho(r) \, = \,
\frac{\rho_s}{\left(\frac{r}{r_s}\right)^\gamma
\left(1+\frac{r}{r_s}\right)^{3-\gamma}} ~,~\,
\end{equation}
where $\rho_s$ is the scale density and $r_s\equiv r_{vir}/c_{vir}(1-\gamma)$ is the scale radius, with $r_{vir}$ the virial
radius of the halo\footnote{The virial radius is usually defined as the range inside
which the average density of DM is some factor of the critical density
$\rho_c$, e.g., $18\pi^2+82x-39x^2$ with $x=\Omega_M(z)-1=-\frac{
\Omega_{\Lambda}}{\Omega_M(1+z)^3+\Omega_{\Lambda}}$ for a $\Lambda$CDM
universe \cite{Bryan:1997dn}.} and $c_{vir}$ the concentration parameter. In this work the
concentration parameter $c_{vir}$ and shape parameter $\gamma$ are left
free, and we normalize the local DM density to be $0.3$ GeV cm$^{-3}$.
Then the virial radius and total halo mass are solved to get self-consistent
values. Given the density profile, the $\gamma$-ray flux along a specific
direction can be written as
\begin{eqnarray}
\phi(E,\psi)&=&C\times W(E)\times J(\psi) \nonumber\\
            &=&\frac{\rho_{\odot}^2R_{\odot}}{4\pi}\times\frac{\langle
\sigma v\rangle}{2m_{\chi}^2}\frac{{\rm d}N}{{\rm d}E}\times
\frac{1}{\rho_{\odot}^2R_{\odot}}\int_{LOS}\rho^2(l){\rm d}l~,~\,
\label{jpsi}
\end{eqnarray}
where the integral is taken along the line-of-sight, $W(E)$ and $J(\psi)$
represent the particle physics factor and the astrophysical factor
respectively. Thus, if the particle physics factor is fixed using the
locally observed $e^+e^-$ fluxes, we can get constraints on the
astrophysical factor, and hence the DM density profile, according to
the $\gamma$-ray flux. For the emission from a diffuse region with
solid angle $\Delta\Omega$, we define the average astrophysical factor as
\begin{equation}
J_{\Delta\Omega}=\frac{1}{\Delta\Omega}\int_{\Delta\Omega}J(\psi){\rm
d}\Omega ~.~\,
\end{equation}

The constraints on the average astrophysical factor $J_{\Delta\Omega}$
for the two models are gathered in Table \ref{table:cross}, in which
$J_{\Delta\Omega}^{\rm max}$ shows the maximum $J$ factor corresponding
to Case I, while $J_{\Delta\Omega}^{1\sigma,2\sigma}$ corresponds to
Case II, at the $68\%$ ($1\sigma$) and $95\%$ ($2\sigma$) confidence levels.
The $\gamma$-ray fluxes of the two cases are shown in the upper panels
of Fig. \ref{gc}.

In the lower panels of Fig. \ref{gc} we show the iso-$J_{\Delta\Omega}$
lines in the $\gamma-c_{vir}$ plane for Case I (left) and Case II (right)
respectively. In this figure we also show the mass condition of
$(1-2) \times 10^{12}$ M$_{\odot}$ of the Milky Way halo.
%For $e^+e^-$ and $\mu^+\mu^-$ final states we only
%consider radiation by internal bremsstrahlung. For the
%$\tau^+\tau^-$ states $\gamma$ rays from the decay of $\tau$ are also
%included. We notice that for the $\tau^+\tau^-$ states the HESS
%constraint is very stringent.
From Fig. \ref{gc} we can see that the NFW profile with $\gamma=1$
(chosen based on N-body simulation in the standard CDM scenario) is
constrained by the HESS data, if the observed cosmic $e^\pm$
excesses are interpreted as DM annihilation. However, if DM is produced
non-thermally the high velocity of the DM particle will make the
DM behave like warm DM and lead to a flat DM
profile which suppresses the $\gamma$-ray flux from the GC.
% Further, the large annihilation cross
%section will account for PAMELA and ATIC and further a flat DM
%profile at the GC will satisfy the HESS observations.

\begin{table}[htb]
\centering
\caption{Parameters of the two scenarios adopted to fit the
ATIC/PPB-BETS/PAMELA or Fermi-LAT/HESS/PAMELA data.}
\begin{tabular}{ccccccc}
\hline \hline
 & channel  & $m_\chi$(GeV) & $\langle\sigma v\rangle$($10^{-23}$cm$^3$ s$^{-1}$) & $J_{\Delta\Omega}^{\rm max}$ & $J_{\Delta\Omega}^{1\sigma}$ & $J_{\Delta\Omega}^{2\sigma}$\\
\hline
Model I & $e^+e^-$ & $620$ & $0.75$ & $300$ & $42$ & $97$\\
Model II & $\mu^+\mu^-$ & $1500$ & $3.6$ & $200$ & $81$ & $111$\\
  \hline
  \hline
\end{tabular}
\label{table:cross}
\end{table}

%%%%%%%%%%%%%%%%%%%%%%%%%%%%%%%%%%%%%%%%%%%%%%%%%%%%%%%%%%%%%%%%%%%

\section{Discussion and Conclusions}

In this paper we have proposed a DM model and studied
aspects of its phenomenology. We have shown that our model can simultaneously
explain the cosmic ray anomalies recently measured by the ATIC,
PPB-BETS and PAMELA experiments or by the Fermi-LAT, HESS and PAMELA
experiments, resolve the small-scale structure problems
of the standard $\Lambda$CDM paradigm, explain the observed
neutrino mass hierarchies, explain the baryon number asymmetry via
non-thermal leptogenesis and suppress the $\gamma$ ray radiation
from the GC.

In this model, DM couples only to leptons. In direct detection experiments
it would show as an ``electromagnetic'' event rather than a nuclear recoil.
Experiments that reject electromagnetic events would thus be ignoring the
signal. However, in the Fermi-LAT/HESS/PAMELA fits, the DM particle couples
mainly to muons, and there being no muons in the target of direct
detection experiments, no significant signal would be expected. In the
ATIC/PPB-BETS/PAMELA fit, the DM couples predominantly to electrons; the
electron recoil energy is of order $m_e v_{\rm DM}^2 \sim 0.1$ eV, and it
would be too small to be detectable in current devices. Alternatively, this
energy could cause fluorescence \cite{Starkman:1995ye}, albeit the
fluorescence cross section would be prohibitively small. Regarding the
annual modulation signal observed by DAMA \cite{dama}, although this
experiment accepts all recoil signals, an estimate of the electron
scattering cross section shows that the present model predicts a cross
section which is about $8$ orders of magnitude smaller than $\sim 1
{\rm pb}$ required to account for the modulation \cite{lepto1}.
Therefore we do not expect a signal in direct detection experiments if the
DM model presented here is realized. In addition, the capture of
DM particles in the Sun or the Earth is also impossible since the DM will not
loose its kinetic energy when scattering with electrons in the Sun.
Therefore we do not expect high energy neutrino signals from the
Sun or the Earth either.

%Our model is based on adding a new DM sector
%with a $U(1)'$ gauge symmetry to the Standard Model and
%introducing  an additional discrete symmetry . The discrete
%symmetry ensures the existence of two degenerate stable DM
%particles. The breaking of the $U(1)'$ symmetry leads to
%the generation of cosmic strings. The decay of cosmic string loops
%is a channel for producing a non-thermal distribution of DM.
%This non-thermal mechanism allows a large annihilation
%cross section to account for the recently observed
%positron/electrion excess by PAMELA and ATIC collaborations cosmic
%while at same time suppress the $\gamma$ ray radiation at the GC
%to meet the constraint by HESS observation.

\section*{Acknowledgments}

We thank Pei-Hong Gu for helpful discussions. We wish to
acknowledge the hospitality of the KITPC under their program
``Connecting Fundamental Theory with Cosmological Observations"
during which the ideas reported here were discussed. This work was
supported in part by the Natural Sciences Foundation of China
(Nos. 10773011, 10821504, 10533010, 10675136), by the Chinese
Academy of Sciences under the grant No. KJCX3-SYW-N2, by the
Cambridge-Mitchell Collaboration in Theoretical Cosmology, and by
the Project of Knowledge Innovation Program (PKIP) of Chinese
Academy of Sciences, Grant No. KJCX2.YW.W10. RB wishes to thank
the Theory Division of the Institute of High Energy Physics (IHEP)
and the CERN Theory Division for hospitality and financial
support. RB is also supported by an NSERC Discovery Grant and by
the Canada Research Chairs Program. PG thanks IHEP for hospitality
and acknowledges support from the NSF Award PHY-0756962.

%%%%%%%%%%%%%%%%%%%%%%%%%%%%%%%%%%%%%%%%%%%%%%%%%%%%%%%%%%%%%%%%%%%%%%%%%%%%

\end{document}